\documentclass[iop]{emulateapj}
\usepackage{color,enumitem,bm,amsmath}
\definecolor{grey}{gray}{0.7}
\newcommand{\funits}{erg~cm$^{-2}$~s$^{-1}$}

\newcommand{\hunits}{$\rm km \ s^{-1} Mpc^{-1}$}
\newcommand{\kmps}{km s$^{-1}$}

\newcommand{\lunits}{erg~s$^{-1}$}

\newcommand{\mstar}{$M_{\ast}$}
\newcommand{\msun}{$M_{\odot}$}
\newcommand{\msuny}{\msun yr$^{-1}$}

\newcommand{\tmone}{$^{-1}$}
\newcommand{\tmtwo}{$^{-2}$}

\newcommand{\lx}{$L_X$}

\newcommand{\lxpoint}{$L_{X\rm, point}$}

\newcommand{\lxgalp}{$L_{X,{\rm gal,p}}$}
\newcommand{\lxgal}{$L_{X,{\rm gal}}$}
\newcommand{\lxlmxb}{$L_{X,{\rm LMXB}}$}

\newcommand{\nh}{$N_{\rm H}$}

\newcommand{\nhgal}{$N_{\rm H}^{\rm gal}$} 
\newcommand{\nhint}{$N_{\rm H}^{\rm int}$}

\newcommand{\x}{X-ray}

\newcommand{\zsun}{$Z_{\odot}$}

\newcommand{\hone}{H{\sc \,i}}
\newcommand{\htwo}{H{\sc \,ii}}

\newcommand{\chandra}{{\it Chandra}}

\newcommand{\spitzer}{{\it Spitzer}}

\newcommand{\swift}{{\it Swift}}

\newcommand{\xspec}{{\sc xspec}}
\newcommand{\wav}{{\sc wavdetect}}

\newcommand{\fr}{Fig.~\ref}
\newcommand{\scr}{Sec.~\ref}
\newcommand{\tr}{Table~\ref}

\newcommand{\exi}{\begin{equation}}
\newcommand{\exo}{\end{equation}}

\newcommand{\aer}[3]{$#1^{#2}_{#3}$} 
\newcommand{\ten}[2]{$#1\times 10^{#2}$} 

\def\spose#1{\hbox to 0pt{#1\hss}} 
\def\approxlt{\mathrel{\spose{\lower 3pt\hbox{$\sim$}}
        \raise 2.0pt\hbox{$<$}}}
\def\approxgt{\mathrel{\spose{\lower 3pt\hbox{$\sim$}}
        \raise 2.0pt\hbox{$>$}}}

\newcommand{\ck}{\checkmark}
\newcommand{\bcred}[1]{{\color{red}{\bf #1}}}
\newcommand{\bcblue}[1]{{\color{blue}{\bf #1}}}
\newcommand{\bcgrey}[1]{{\color{grey}{\bf #1}}}

\newcommand{\pba}{$P_{b,1}$}
\newcommand{\pbb}{$P_{b,2}$}
\newcommand{\pcg}{$P_{\rm CG}$}

\slugcomment{Accepted by ApJ}

\shorttitle{\chandra\ X-ray Binaries in Compact Groups}
\shortauthors{Tzanavaris et al.}

\begin{document}

\title{Exploring x-ray binary populations in compact group galaxies with Chandra}

\author{
P.~Tzanavaris\altaffilmark{1,2,3},
A.~E.~Hornschemeier\altaffilmark{1}, 
S.~C.~Gallagher\altaffilmark{4}, 
L.~Lenki{\'c}\altaffilmark{4},
T.~D.~Desjardins\altaffilmark{5},
L.~M.~Walker\altaffilmark{6},
K.~E.~Johnson\altaffilmark{7},
J.~S.~Mulchaey\altaffilmark{8}
}

\altaffiltext{1}{Laboratory for X-ray Astrophysics, 
NASA/Goddard Spaceflight Center, Mail Code 662, Greenbelt, Maryland, 20771, USA}
\altaffiltext{2}{CRESST, University of Maryland Baltimore County, 1000 Hilltop Circle, Baltimore, MD 21250, USA}
\altaffiltext{3}{Department of Physics and Astronomy,
The Johns Hopkins University, Baltimore, MD 21218, USA}
\altaffiltext{4}{Department of Physics and Astronomy \& Centre for Planetary and Space Exploration, The University of Western Ontario, London, ON N6A 3K7, Canada}
\altaffiltext{5}{Department of Physics and Astronomy, 
177 Chem.-Phys. Building, University of Kentucky,
505 Rose Street, Lexington KY 40506-0055202, USA}
\altaffiltext{6}{Steward Observatory, University of Arizona, Tucson, AZ 85721, USA}
\altaffiltext{7}{Department of Astronomy, University of Virginia, P.O. Box 400325, Charlottesville, VA 22904-4325, USA}
\altaffiltext{8}{The Observatories of the Carnegie Institution for Science, Pasadena, CA 91101, USA}

\begin{abstract}
We obtain total galaxy \x\ luminosities, \lx,
originating from individually detected point sources in
a sample of 47 galaxies in 15 compact groups of galaxies (CGs).
For the great majority of our galaxies, we find that
the detected point sources most likely are local to their
associated galaxy, and are thus extragalactic \x\ binaries (XRBs) or nuclear
active galactic nuclei (AGNs). For spiral and irregular galaxies, 
we find that, after accounting for AGNs and nuclear sources,
most CG galaxies are either within the $\pm1\sigma$
scatter of the \citet{mineo2012} \lx\ - star formation rate (SFR) correlation or
have higher \lx\ than predicted by this correlation for their SFR. 
We discuss how these ``excesses'' may be due to low metallicities and high
interaction levels.
For elliptical and S0 galaxies, after accounting for AGNs
and nuclear sources, most CG galaxies are consistent with the
\citet{boroson2011}
\lx\ - stellar mass correlation for low-mass XRBs, with larger scatter, likely due to
residual effects such as AGN activity or hot gas.
Assuming non-nuclear sources
are low- or high-mass XRBs, we use appropriate XRB luminosity
functions to estimate the probability that
stochastic effects can lead to such extreme \lx\ values.
We find that, although stochastic effects do not in general appear to be 
important, for some galaxies there is a significant probability that
high \lx\ values can be observed due to strong XRB variability.
\end{abstract}

\vspace{1cm}
\keywords{X-rays: galaxies -- X-rays: binaries}

\section{Introduction}
Most galaxies in the local Universe are found within poor
groups \nocite{karachentsev2005,mulchaey2000} (Mulchaey 2000 and
references therein; Karachentsev et al. 2005).  
Studying the
physical processes in these systems is thus fundamental to
understanding galaxy formation and evolution, which remains one of the
key goals of extragalactic astrophysics. 

Compact galaxy groups constitute an extreme class of such small galaxy
systems. The separations of galaxies in these systems are
typically of the order of just a few galaxy radii, with
median projected separations $\sim 40h^{-1}$ kpc\footnote{$h\equiv H_0/100$, where $H_0$ is the Hubble constant at redshift 0.}); velocity
dispersions are low, with a radial median $\sim 200$ \kmps,
while galaxy number
densities are high \citep[up to $10^8 h^2$
  Mpc$^{-2}$,][]{hickson1982,hickson1992}. Thus CGs are environments where
galaxy interactions are prevalent. Such interactions give rise to
tidal tails and favor collapse of giant molecular clouds, thus
affecting star formation and its evolution \citep[e.g.][]{mihos1996}.

Star formation in CG galaxies appears to be
accelerated when compared to galaxies
that do not reside in such an environment.
This in turn likely leads to fast consumption of the available gas
and abrupt quenching.
The observational signatures of this are ``gaps'' both in
mid-infrared colors \citep{johnson2007,walker2010,walker2012}
as well as mass-normalized star formation rates (specific
SFRs, sSFRs, Tzanavaris et al. 2010, Lenkic et al. 2015, in prep.).
The latter works further suggest
that the more strongly star-forming galaxies are preferentially
hosted by CGs which are richer in \hone\ gas, and vice versa
for galaxies that show low levels of star-forming activity.

The interaction-prone CG environment may also affect the type and
level of activity occurring in the {\it nuclei} of CG galaxies. For
instance, simulations have shown that major and minor galaxy mergers
can often bring about inflow of rotationally supported gas from the
outer parts of a galaxy to fuel central supermassive black hole
activity via accretion \citep[\lq\lq AGNs\rq\rq, e.g.][]{hopkins2010},
and this is corroborated by observations
\citep[e.g.][]{kartaltepe2010}.
Several works using a variety of
multi-wavelength data and methods have shown that, although CGs do
host substantial {\it numbers} of AGNs, the {\it level} of this
activity does not, in general, appear to be high \citep[see][hereafter
  T14, and references therein]{tzanavaris2014}.

\x\ emission provides an alternative path to probing star formation in
the CG environment. 
Broadly speaking,
for galaxies without strong \x\ AGNs,
\x\ emission is dominated by X-ray
binary stars (XRBs), where a compact primary, either a neutron star or
black hole, accretes matter from either a low-mass (LMXBs) or a high-mass 
(HMXBs) donor secondary
in early- and late-type galaxies, respectively
\citep{kim1992}.  Thanks to the \chandra\ \x\ telescope's sub-arcsecond
resolution, it is now possible to detect such individual \x\ point
sources in nearby galaxies \citep[see, e.g.][and references
  therein]{fabbiano2006}. If these sources are offset from a galaxy's
nucleus, these are most likely either LMXBs or HMXBs. Combined with
the estimated high binarity fraction in stellar populations, the study
of these sources can provide key information for galaxies as a whole.
The situation is less clear for nuclear sources because
their \x\ emission may come from an AGN, star formation \citep[e.g.][]{rossa2006}, or both types of activity.

In actively star-forming (morphologically spiral/irregular) galaxies,
the SFR is traced by populations of massive, luminous, and short lived
($\lesssim 100$~Myr) OB associations.  Some of these stars will be
HMXB secondaries: Indeed several studies of late-type galaxies
have established strong
correlations between galaxy-wide \x\ luminosity and total SFR
\citep[e.g.][]{bauer2002,grimm2003,ranalli2003,gilfanov2004a,persic2004,hornschemeier2005,persic2007,lehmer2008,lehmer2010,mineo2012}.
On the other hand, in quiescent (morphologically elliptical/S0)
galaxies, longer-lived ($\gtrsim 1$~Gyr) low-mass stars trace the
total stellar mass accumulated over a galaxy's lifetime. Such stars may be
LMXB donors, which explains the observed correlations between an early-type
galaxy's \x\ luminosity and total
stellar mass, \mstar\ \citep[e.g.][]{fabbiano2002,gilfanov2004b,kim2004,lehmer2010,boroson2011}.

Clearly, this is an oversimplified picture. 
Many galaxies will have a mix of HMXB and LMXB populations, such
as ellipticals with residual star formation or bulge-dominated spirals.
In general detailed modeling is therefore required to
understand the contributions of different sub-populations to the total
X-ray point-source luminosity 
both for late- and for early-type galaxies \citep[see][for a population synthesis based analysis of a combined LMXB and HMXB population]{luo2012}.

\citet[][hereafter M12]{mineo2012} study 29 nearby ($<20$~Mpc) galaxies\footnote{With the exception of one galaxy at $\sim 123$~Mpc.}.
This \lq\lq primary\rq\rq\ sample is selected based on
several criteria, such as morphologies, sSFR
threshold, identification and exclusion of bulge sources, and
sensitivity to ensure that their systems are late-types dominated by
resolved HMXB emission. They also perform detailed corrections for
incompleteness at the low luminosity end. 
The galaxies in this primary sample 
(up to $\lesssim 20$~\msuny)
include the moderate SFR range,
which is representative of CG galaxies,
and have a variety of late-type
morphologies, providing a control sample for CG late-type systems.

In their study of 17 nearby ($< 60$~Mpc) luminous IR galaxies,
\cite{lehmer2010}
take advantage both of high-resolution
imaging with \chandra\ and spectral information to disentangle the
contributions of AGNs, XRBs and hot gas. They parameterize and obtain
best fits for the 
combined
contributions of LMXBs and HMXBs as
functions of \mstar\ and SFR, respectively. Their sample was
selected by proximity and low column density, in order to maximize
separability of the various contributions to \x\ emission.
\citet{lehmer2010}
perform fits as functions of sSFR; however their sSFR regime
does not cover the substantial low-sSFR CG regime. 
A direct comparison of CGs with these fits may then
be not particularly useful.

\citet[][hereafter B11]{boroson2011} study exclusively
\x\ emission from 30 early-type galaxies observed with \chandra\ to
a depth that allows the detection of individual LMXBs with 
\lx~$> 10^{38}$~\lunits. In their analysis, they
estimate in detail the separate contributions to the total \x\
luminosity due to active binaries (ABs), cataclysmic variables (CVs),
hot gas and LMXBs. They thus obtain a scaling relation between
\lxlmxb\ and $K$-band luminosity, $L_K$, which traces old
stellar populations and, hence, stellar mass. 

The B11 sample is well-matched in stellar mass to the CG sample
and is thus suitable as a comparison sample for
the LMXB emission from CG early-type galaxies.

We have already studied {\it diffuse} \x\ emission in CGs in two recent
publications.  In \citet{desjardins2013} and \citet{desjardins2014} we detect diffuse \x\ emission in a majority of CGs.  In dynamically
unevolved systems, characterized by low masses and velocity dispersions,
we find that
this emission originates in processes local to the galaxies
(star formation, AGNs or tidal tails) and appears directly linked
to individual galaxies rather than a diffuse intragroup medium
(IGM). 

In this paper we use \chandra\ archival data to 
detect individual \x\ point sources in a sample of
47 CG galaxies. 
The aim of the paper is to assess
whether, and to what extent, point-source based
galaxy luminosities for CG galaxies
agree with the \lx\ - SFR and
\lx\ - \mstar\ correlations for late- and early-type galaxies, respectively.
The structure of the paper is as follows:
 Section 2 describes the CG data sample.
Section 3 discusses X-ray data and analysis, with emphasis on point source
and galaxy-wide luminosities, 
as well as stochastic effects, and local and background AGN contamination.
Section 4 is a brief description of the SFR and \mstar\ obtained
via non-\x\ work, which is presented in a separate forthcoming
publication (Lenkic et al. 2015, L15, in prep.).
Section 5 presents our CG results within the context
of \x\ luminosity scaling relations.
We discuss these results in Section 6, and
summarize in Section 7.
We use $\Omega_M = 0.3$, $\Omega_{\Lambda} = 0.7$, and
$H_0 = 70$ \hunits\ throughout.
\begin{deluxetable*}{cccc cccc c}
\tablecolumns{9}
\tablewidth{0pc} 
\tablecaption{Compact Group Sample \label{tab-sample}}
\tablehead{ 
\colhead{ID}
&\colhead{$z$}
&\colhead{$v$}
&\colhead{$D_L$}
&\colhead{\hone\ type}
&\multicolumn{4}{c}{\chandra\ \x}
\\
&
&(\kmps)
&(Mpc)
&
&obs.~IDs
&ACIS
&ks
&refs
\\
\colhead{(1)}
&\colhead{(2)}
&\colhead{(3)}
&\colhead{(4)}
&\colhead{(5)}
&\colhead{(6)}
&\colhead{(7)}
&\colhead{(8)}
&\colhead{(9)}
}
\startdata
  HCG~7  & 0.0141   &    3885 &  56.0 &    II &  8171, 9588        & S & 36.3  & T14 \\
  HCG~16 & 0.0132   &    3706 &  53.4 &    II &  923               & S & 12.7  & T14 \\
  HCG~22 & 0.0090   &    2522 &  36.3 &    II &  8172              & S & 32.2  & T14 \\
  HCG~31 & 0.0135   &    4026 &  58.1 &     I &  9405              & S & 36.0  & T14 \\
  HCG~37 & 0.0223   &    6940 & 101.0 &   III &  5789              & S & 17.9  & D14 \\
  HCG~40 & 0.0223   &    7026 & 102.0 &    II &  5788, 6203        & S & 48.3  & D14 \\
  HCG~42 & 0.0133   &    4332 &  62.6 &    II &  3215              & S & 32.1  & T14 \\
  HCG~59 & 0.0135   &    4392 &  63.5 &   III &  9406              & S & 38.9  & T14 \\
  HCG~62 & 0.0137   &    4443 &  68.7 &   III &  921, 10462, 10874 & S & 169.2 & T14 \\
  HCG~79 & 0.0145   &    4439 &  64.1 &    II &  11261             & S & 69.2  & D14 \\
  HCG~90 & 0.0088   &    2364 &  34.0 &   III &  905               & I & 50.2  & T14 \\
  HCG~92 & 0.0215   &    6119 &  88.8 &    II &  789, 7924         & S & 114.4 & T14 \\
  HCG~97 & 0.0218   &    6174 &  89.6 &   III &  4988              & S & 57.4  & D14 \\
  HCG~100& 0.0178   &    4976 &  72.0 &    II &  6978, 8491        & I & 45.6  & D14 \\
  RSCG~17& 0.0190   &    5415 &  78.4 &    II &  2223              & S & 30.4  & D14 
\enddata
\tablecomments{Columns are: (1) Compact group name; (2) redshift; (3) velocity from NASA Extragalactic Database (NED) relative to
the cosmic microwave background velocity \citep{fixsen1996}; (4) luminosity distance from NED; (5) group evolutionary \hone\ type as measured and defined by \citet{johnson2007}: (I) \hone\ rich ($\log M_{\rm H I}/\log M_{\rm dyn} > 0.9$); (II) intermediate ($\log M_{\rm H I}/\log M_{\rm dyn} = 0.8-0.9$); and (III) \hone\ poor ($\log M_{\rm H I}/\log M_{\rm dyn}< 0.8$); (6) \x\ observations IDs; (7) \chandra-ACIS array; (8) exposure time; (9) for further details and references on the \x\ data see these papers: D14 \citep{desjardins2014}, T14 \citep{tzanavaris2014}.
}
\end{deluxetable*}

\section{Sample selection}
In this paper we study a sample of 15 compact groups observed with
\chandra/ACIS, \swift/UVOT and \spitzer/IRAC-MIPS for which archival
data exist, allowing us to obtain SFRs, stellar masses, sSFRs and
\x\ fluxes and luminosities. \tr{tab-sample} shows the group sample, including
redshifts, luminosity distances and group evolutionary types. Allowing
for the fact that some galaxies do not fall in the field of view of
all three instruments, the total number of CG galaxies analyzed is 47.

The majority of the CGs analyzed here are from the
catalog of 92 CGs that have been spectroscopically
confirmed by \citet{hickson1992}.  
As in \citet{tzanavaris2010} and \citet{tzanavaris2014}
the selection criteria for this sample aimed to allow
observations with a range of telescopes and instruments, both
space and ground-based.
We thus demanded that a CG contain at least three giant
galaxies with ``accordant'' redshifts  within 1000 \kmps\ of the
group's mean redshift, be closer than $\sim 4500$ \kmps, and
span less than $\sim 8$\arcmin\ on the sky.  
In spite of concerns about
selection biases in the Hickson catalog
\citep{mamon1994,ribeiro1998}, it has been shown
that several characteristic properties of HCG galaxies are
comparable to median galaxy values \citep[e.g. surface brightness and
linear or angular diameters,][T14]{lee2004}). 
One of
our CGs comes from the catalog of Redshift Survey Compact Groups
\citep[RSCGs,][]{barton1996}. This RSCG is part of a larger sample
selected by \citet{walker2012} to have properties similar to those of
HCGs.

Details on the \swift\ and \spitzer\ observations and data for systems
in this sample can be found in \citet{tzanavaris2010} and L15. For
\chandra/ACIS observations we refer the reader to T14 and
\citet{desjardins2013,desjardins2014}.

\section{X-ray data analysis}
\subsection{Point source detection and photometry}
The galaxy groups were observed with \chandra 's Advanced CCD Imaging
Spectrometer (ACIS), either with the back-illuminated
S3 CCD or the ACIS-I array. 
The data reduction and analysis, as well as point source detection,
for CGs labelled T14 in Table 1, Column 6, was done as described 
in T14. For the rest of the CGs,
the data reduction and analysis
was carried out using the CIAO
4.5\footnote{http://cxc.harvard.edu/ciao} \chandra\ \x\ Center data
reduction and analysis suite.  Standard
aspect solution and grade filtering was used, leading
to the production of a final
level 2 events
file. In the case of multiple observations, observations
were reprojected and individual exposure maps created for each
observation, before obtaining summed counts and exposure map images
({\tt merge\_obs} in CIAO).

\chandra\ has an unprecedented angular resolution that makes it
the instrument of choice for \x\ point source detection. However,
this resolution (or, equivalently, the \chandra\ point spread function,
PSF) varies across the ACIS CCDs from less than one arcsecond
near the aim point to several arcseconds near the CCD edges. 
Information about this variation must be taken into account
before reliable point source detection and characterization
can be carried out. For each observation we thus
constructed PSF maps ({\tt mkpsfmap} in CIAO) and in the case
of multiple observations we also produced a final exposure-weighted
mean PSF map.

The PSF map information was used as input at the next stage
where we carried out point source detection
by correlating the data with wavelet functions 
\citep[\wav\ in CIAO,][]{freeman2002}
in the 
0.5--8.0 keV energy band. 
Below 0.5 keV the response is not well calibrated, while
above 8.0 keV both the mirror effective area drops and the
particle background rises.
The false-probability threshold ($10^{-6}$ and $2.4\times 10^{-7}$, for S3 and 
I0-I3, respectively) was chosen to match
the inverse of the
size of the chip field, i.e. 
1024$\times$1024 pixels for chip S3 and 2048$\times$2048 pixels for
chips I0-I3.
To cover a broad range of source sizes, and
account for the varying point spread function (PSF) size across
the ACIS CCD, wavelet scales of 1, 1.414, 2, 2.828, 4,
5.657 and 8.0 pixels were used.
This stage produced catalogs of {\it candidate} point sources.

Each candidate point source has an elliptical source-region assigned to it, 
for which the local PSF size has been taken into account. 
We multiplied the ellipse axes by a factor of five and excluded the source
region to construct background annuli. These were also
checked visually and any non-background emission was further
excluded. Net counts were calculated by
subtracting area-normalized background counts from source counts.
Sources with zero or negative net 
counts were discarded as non-detections.
Poisson $\pm 1\sigma$ errors on net counts were estimated by
using the method of \citet{gehrels1986}. 
Sources were also considered as non-detections and discarded
if the net counts, after subtraction of
the lower $2\sigma$ error, were negative.
Finally, for detected sources we extracted spectra and associated ancillary response
and response matrix files (ARF and RMF, {\tt specextract} in CIAO).

\subsection{\x\ luminosities}
In this paper we calculate and/or refer to several different types of
\x\ luminosities, a summary of which is presented in
\tr{tab-lxtypes}. 
We further discuss these below.

\begin{deluxetable}{ll}
\tablecolumns{2}
\tablewidth{0pc} 
\tablecaption{\x\ luminosities used in this paper\label{tab-lxtypes}}
\tablehead{ 
\colhead{Name}
&\colhead{Description}
}
\startdata
\lxpoint\          & \x\ luminosity of a single detected point source. \\
\lx\               & Galaxy wide \x\ luminosity from sum of \lxpoint\ \\
                   & values within a galaxy region.   \\
\lxgal\            & Galaxy wide \x\ luminosity based on total counts  \\
                   & within a galaxy region.                           
\enddata
\end{deluxetable}

\subsubsection{Point sources}
Because the majority of point sources
have few net counts, reliable spectral fitting
cannot be carried out.
Instead, as in T14, we estimate \x\ spectral shapes, fluxes and luminosities 
following
\citet{gallagher2005}.
If $H$ and $S$ are net counts in the hard ($2.0-8.0$~keV)
and soft ($0.5-2.0$~keV)
bands, respectively, the hardness ratio is given by HR $\equiv (H-S) /(H+S)$. 
For each \x\ point source we compare the observed
hardness ratio to that obtained from simulated spectra, and
estimate the power-law index $\Gamma$ (where the photon flux is
$f_E \propto E^{-\Gamma}$
photon cm\tmtwo\ s\tmone\ keV\tmone) and corresponding \x\ flux and
luminosity. 

We produce a set of simulated spectra with {\sc xspec} \citep{arnaud1996}, version 12.8.1g, for each point source, using
the corresponding Galactic column
density, \nhgal, as well as the ARFs and RMFs from the spectral
extraction.
Each set of spectra uses an
absorbed power law model ({\tt tbabs*po} in {\sc XSPEC}) with
$\Gamma$ varying between $-1$ and $+4$, giving simulated
count rates in the soft and hard band, from which simulated HR values
follow. By comparing the observed HR from net counts for each observed detected source within the source extraction region with the simulated HR values, 
we estimate the best $\Gamma$
value, and, thus, corresponding flux and point-source luminosity, \lxpoint, for
a given \x\ point source.

As the assumed model is simple and assumes no intrinsic absorption (\nhint~=~0), 
some $\Gamma$ values may in fact be erroneous. However, there is
a degeneracy between $\Gamma$ and
\nh, so that such errors should not affect luminosity values, which is
our ultimate goal.

\subsubsection{Galaxy wide}\label{sec-wide}
For each galaxy, we calculate the total \x\ luminosity that is due to
point sources, \lx, by first identifying detected point sources that
fall within the boundaries of a given galaxy, and then simply summing
individual \lxpoint\ values, i.e.

\exi 
L_X = \sum_{\rm gal} L_{X,{\rm point}} \, \, .
\exo

Galaxy boundaries are Kron apertures \citep{kron1980} obtained by
running
SEXtractor \citep{bertin1996} on the \spitzer/IRAC 3.6\micron\ images, 
which are the best tracers of stellar mass among our multiwavelength data.
However, it is not always possible to define a single aperture for a galaxy,
e.g. in cases of strongly interacting systems. We use a single, combined
region for such cases (see L15 for more details). Thus,
in this paper the term \lq\lq galaxy\rq\rq\ or \lq\lq galaxy
region\rq\rq\ refers to these Kron apertures, which in some cases include more
than one letter-designated galaxy.

\begin{figure}
\hspace{-2cm}
\includegraphics[scale=0.75,clip=true]{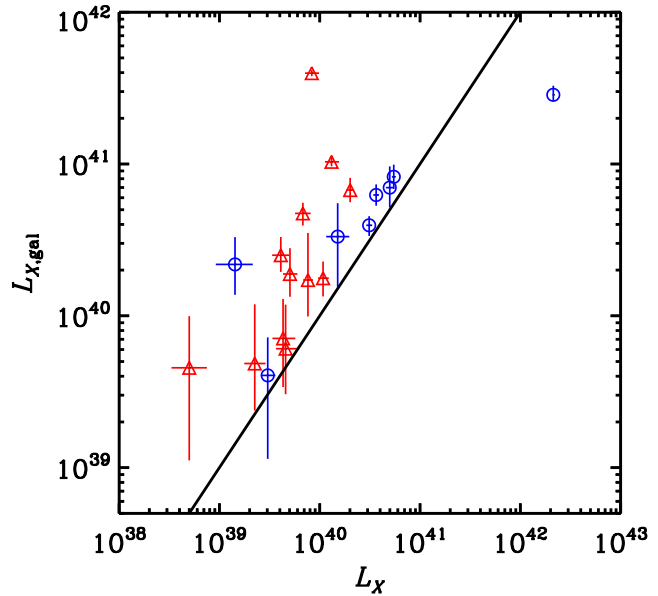}
\caption{``Galaxy-wide'' luminosity, \lxgal, against total point-source
  luminosity, \lxgalp, for a sub-sample of our CG galaxies.  Blue
  circles are late types and red triangles early types. The solid line
  is the equality locus. The source at top right is HCG~92C for which
  the nuclear emission has been subtracted in the \lxgal\ estimate. As
  expected, \lxgal\ overestimates \lx.  }
\label{fig_lxlx}
\end{figure}

Some studies in the literature report total galaxy \x\ luminosities
calculated from the total \x\ emission within galaxy regions, rather than
\x\ emission due only to point sources.
If individually detected point sources are not used, a calculated
galaxy-wide \x\ luminosity will not necessarily reflect the level of
emission due to LMXBs and HMXBs. For the purpose of comparison with
such work, we also calculate
galaxy-wide luminosities by extracting a single source and background
\x\ spectrum for each Kron aperture associated with a CG galaxy. We
use \xspec\ to fit each spectrum with a power law+thermal plasma,
taking Galactic column density into account. For further details, we
refer the reader to Desjardins et al. (2015, in prep.).
In this paper, we label luminosities derived in this fashion \lxgal,
to distinguish them from the total point source luminosity inside
a galaxy region, \lx. As a sanity check, we compare
the two types of luminosities in \fr{fig_lxlx}, where we plot
\lxgal\ against \lx\ in the 0.5 -- 8.0~keV band. As expected, galaxy-wide
luminosities from total \x\ emission
are consistently higher, with the exception of
HCG~92C where the central AGN has been excised by Desjardins et al. (2015).
We only show galaxies that are truly single systems (i.e.~each
region corresponds to a single letter-designated galaxy), but 
including multiple-galaxy systems doesn't change the result.

\subsubsection{Nuclear Point Sources and Local AGN Contamination}
\x\ point sources located at galaxy centers represent an additional
complication for this work. Such sources may be due either to
nuclear star clusters or an AGN. If the \x\ luminosity is high
($\gtrsim 10^{41}$~\lunits) it is highly probable that one is
dealing with an AGN. Otherwise, as 
discussed at length 
in T14, a variety of multi-wavelength
diagnostics may be used to identify the nature of the source.
Since we are interested in galaxy-wide \x\ luminosities that originate in XRBs,
we identify all nuclear sources 
either by using the earlier results of T14
or, for galaxies not analyzed in that work,
by cross-correlating the 
\chandra/ACIS and \swift/UVOT data. We find that 36 out of a total
of 47 galaxies (77\%) have a nuclear source. Out of these, 
multiwavelength work has already classified those in
the late-type galaxies HCG~16B and 92C, as well as the
early-type HCG~22A as AGNs. HCG~16A is classified as
LINER or weak AGN by T14. We also tentatively classify the nuclei
of RSCGs 17A and 17B as likely AGNs, since the \x\ luminosity is higher
than the threshold above. For the remaining galaxies
the multiwavelength classification is either non-AGN or unknown.
We note that for
12 galaxies (25\%) their single \x\ point source is also
nuclear. 
For all galaxies with nuclear \x\ sources we give the
fraction of the total luminosity that originates in the nuclear 
source in column 7 of \tr{tab-results}. 

Overall, we may conservatively assume that a galaxy that fulfills
any of three criteria may suffer some AGN contamination: \lx$\ge 10^{41}$~\lunits, a single nuclear \x\ source or independent AGN classification.
There are 15 CG galaxies ($\sim32\%$) that fulfill
at least one of these criteria: Five late-types and ten early-types.
These are indicated by red circles in \fr{fig_mineo}
and \fr{fig_bor}.

\subsubsection{\x\ luminosity error}
The error bars associated with \x\ luminosity data
points in our figures represent the observational error due to Poisson
statistics.  However, \citet{gilfanov2004c} have shown that in the low
SFR regime with small numbers of individual \x\ point sources,
stochastic effects can be important, leading to an \x\ luminosity
excess with respect to the \lx\ - SFR correlation. Since in general
the galaxies in our sample have low SFRs and contain few point sources, we
investigate stochastic effects as follows.

For each galaxy and associated luminosity limit (\scr{sec-lims}) we
calculate the total number of point sources that we expect to detect,
$<N>$, by integrating the XRB \x\ luminosity functions (XLFs) of
\citet{gilfanov2004b} and M12 for early- and late-type
galaxies, respectively. For each $<N>$ we carry out 10,000 Monte Carlo (MC)
realizations by drawing a number $N$ from a Poisson distribution with
mean $<N>$. The normalized XLF is a probability distribution, which we
use to randomly assign an \lxpoint\ value to each of the $N$ sources,
and then calculate an \lx\ value for each galaxy and realization.
The fraction of realizations with \lx\ values equal to or greater
than the observationally obtained \lx\ values for each galaxy is
an estimate of the probability, \pba, that one might obtain our
observed \lx\ values stochastically.

Additionally, both HMXBs and LMXBs may be highly variable, which would
exacerbate the issue of small number statistics. The range of possible
\x\ luminosities depends on individual characteristics of each XRB
system such as accretor and donor type, and transient or persistent
emission. On the other hand, it has been suggested that the XRB
XLFs are robust against variability \citep[][and references
  therein]{fabbiano2006}. As we do not have specific information for the
particular XRB types that we observe, we treat each initial \lxpoint\ value in
each of our MC realizations, effectively drawn from the corresponding
XLF, as a mean value, and assume a variability-driven uncertainty of
an order of magnitude above and below this value. We then randomly
draw a final \lxpoint\ value for each source before calculating
\lx\ for each realization and computing the probability, \pbb,
that one might obtain our observed \lx\ values stochastically.

The probabilities \pba\ and \pbb\ are shown in columns 11 and 12 of
\tr{tab-results}. The \pba\ values are generally low, with only
three out of 47 galaxies ($\sim 6\%$) having \pba~$\ge 0.05$.
Thus, if variability is neglected, our \lx\ values are stochastically
improbable. The three galaxies are the late types HCGs~22C, 40BCE and
59A, which agree closely with the M12 \lx-SFR correlation 
(\scr{sec-lxsfr}, \tr{tab-latecorrel} and \fr{fig_mineo}). This
is consistent with the fact that the \pba\ estimate assumes
the M12 XLF for late type galaxies. As one might expect, \pbb\ values
are often higher. About half of all galaxies (26 or $\sim 55\%$)
still have \pbb~$<0.05$; 15 galaxies ($\sim 32\%$) have 
0.05~$\le$~\pbb~$< 0.20$; and 6 galaxies ($\sim 13\%$) have
\pbb~$\ge 0.20$. \pba\ and \pbb\ are further discussed in 
\scr{sec-disc}.

\subsection{Flux Limits and Background AGN Contamination}\label{sec-lims}
Without spectroscopic redshifts, it is possible
that some of the \x\ point sources within CG galaxies
are in fact background AGNs. In our case,
the problem is exacerbated by the small number of sources detected
per galaxy. Following T14, we estimate the level of background AGN contamination as follows.

We first estimate flux and luminosity limits for our sources.  
According to \citep{broos2010}, the binomial probability, $P_B$ for a source to be spurious
is given by the binomial function
\exi P_B = f_b ( C_s;
C_s+C_b, (1+A_b / A_s)^{-1}) \ , \label{equ:binpro} 
\exo 
where $C_s$ and $C_b$ are the number of counts observed in the source
and background region in a specific energy band, and $A_s$, $A_b$ are the
areas of the source and background regions.  Thus, 
if a source
has a value of $P_B$ {\it less} than some given threshold value, it
is considered a detection. We choose the threshold $P_B = 0.004$
determined by \citet{xue2011} together with local-background count numbers
for each detected source to establish a detectability limit in terms
of counts, fluxes and luminosities in the $0.5-8.0$~keV band at the
specific CCD location of each source. 
Because the calculation of background counts for each source has
taken into account the correct local size of the \chandra\ PSF,
the calculated detection limits are specific to each source.
In our case, $C_b$ is simply the measured background counts for each source.
Starting with $C_s =0$, corresponding to $P_B >> 0.004$ and
a clearly spurious source, we increment $C_s$ until a value
$C_{s,{\rm lim}}$ is obtained, corresponding to the chosen
$P_B$ threshold of 0.004.  Converting the estimated $C_{s,{\rm lim}}$ value
to a flux by assuming a power law spectrum with $\Gamma=1.4$
\citep{hickox2006,steffen2007} and the Galactic $N_H$ value for the host
galaxy in {\sc pimms} \citep{mukai1993} produces a detection flux
limit for the location of a source. Where there are several
sources per galaxy we take the lowest source-specific
flux limit as the nominal flux limit for the galaxy. The flux and
corresponding luminosity detection limits thus obtained are shown in columns 
8 and 9 of \tr{tab-results}.

\begin{deluxetable*}{cccc cc cccccc}
\tablecolumns{12}
\tablewidth{0pc} 
\tablecaption{CG Galaxy Results \label{tab-results}}
\tablehead{ 
\colhead{ID}
&\colhead{Morpho}
&\colhead{SFR}
&\colhead{\mstar}
&\colhead{$L_X$}
&\colhead{Number of}
&\colhead{central}
&\colhead{\bm{$f_{\rm lim}$}}
&\colhead{\bm{$L_{X, {\rm lim}}$}}
&\colhead{\bm{$P_{\rm CG}$}}
&\colhead{\bm{\pba}}
&\colhead{\bm{\pbb}}
\\
\colhead{}
&\colhead{logy}
&\colhead{\msuny}
&\colhead{$10^{10}$\msun}
&\colhead{$10^{40}$\lunits}
&\colhead{sources}
&\colhead{\bm{\lx}}
&\colhead{$10^{-15}$~erg}
&\colhead{$10^{38}$\lunits}
&\colhead{}
&\colhead{}
&\colhead{}
\\
\colhead{}
&\colhead{}
&\colhead{}
&\colhead{}
&\colhead{}
&\colhead{}
&\colhead{fraction}
&\colhead{cm$^{-2}$~s$^{-1}$}
&\colhead{}
&\colhead{}
&\colhead{}
&\colhead{}
\\
\colhead{(1)}
&\colhead{(2)}
&\colhead{(3)}
&\colhead{(4)}
&\colhead{(5)}
&\colhead{(6)}
&\colhead{(7)}
&\colhead{(8)}
&\colhead{(9)}
&\colhead{(10)}
&\colhead{(11)}
&\colhead{(12)}
} 
\startdata
     H~7A &     S  & $3.41\pm0.45$  &  $9.17\pm0.04$  & \aer{3.65}{+0.25}{-0.23}   & 2   & 0.27   &    2.2 &    8.3 &    0.6 & 0.01 & 0.17    \\
     H~7B &    S0  & $0.23\pm0.03$  &  $4.65\pm0.05$  & \aer{0.22}{+0.06}{-0.05}   & 2   & 0.45   &    1.0 &    3.6 &    0.5 & 0.00 & 0.28    \\
     H~7C &     S  & $2.00\pm0.23$  &  $4.51\pm0.04$  & \aer{5.47}{+0.25}{-0.23}   & 6   & 0.02   &    1.6 &    6.0 &    0.9 & 0.00 & 0.05    \\
     H~7D &    S0  & $0.43\pm0.05$  &  $1.37\pm0.06$  & \aer{0.76}{+0.09}{-0.08}   & 2   & \ldots &    1.0 &    3.6 &    0.6 & 0.00 & 0.01    \\
    H~16A &     S   & $4.65\pm0.60$  &  $10.50\pm0.04$  & \aer{4.99}{+0.46}{-0.42} & 2   & 0.94   &    7.3 &   24.8 &    0.9 & 0.01 & 0.15    \\
    H~16B &     S   & $0.48\pm0.06$  &  $6.41\pm0.04$  & \aer{22.00}{+1.70}{-1.60} & 1   & 1.00   &    7.3 &   24.8 &    0.9 & 0.00 & 0.00    \\
    H~16C &    S0   & $13.98\pm1.85$  &  $6.74\pm0.03$  & \aer{1.31}{+0.25}{-0.18} & 4   & 0.40   &    4.5 &   15.5 &    0.9 & 0.00 & 0.00    \\
    H~16D &    S0   & $16.73\pm2.26$  &  $3.79\pm0.02$  & \aer{0.68}{+0.13}{-0.11} & 1   & 1.00   &    5.4 &   18.6 &    0.8 & 0.00 & 0.00    \\
    H~22A &     E   & $0.36\pm0.04$  &  $9.14\pm0.02$  & \aer{1.08}{+0.15}{-0.12}  & 12  & 0.05   &    1.1 &    1.7 &    0.8 & 0.00 & 0.35    \\
    H~22B &    S0   & $0.06\pm0.01$  &  $0.76\pm0.03$  & \aer{0.05}{+0.02}{-0.02}  & 1   & 1.00   &    1.4 &    2.3 &    0.7 & 0.04 & 0.11    \\
    H~22C &     S   & $0.55\pm0.06$  &  $0.80\pm0.02$  & \aer{0.30}{+0.06}{-0.04}  & 3   & \ldots &    1.4 &    2.3 &    0.7 & 0.08 & 0.17    \\
  H~31ACE &     S   & $9.93\pm1.17$  &  $2.27\pm0.04$  & \aer{9.60}{+0.44}{-0.39}  & 9   & 0.23   &    1.0 &    3.9 &    0.9 & 0.01 & 0.28    \\
    H~31B &     S   & $1.06\pm0.11$  &  $0.36\pm0.20$  & \aer{2.11}{+0.22}{-0.19}  & 3   & \ldots &    1.3 &    5.2 &    1.0 & 0.01 & 0.09    \\
    H~31G &     S   & $1.83\pm0.20$  &  $0.71\pm0.10$  & \aer{3.11}{+0.23}{-0.20}  & 5   & 0.31   &    1.3 &    5.2 &    1.0 & 0.01 & 0.08    \\
    H~31Q &     Im  & $0.13\pm0.02$  &  $0.20\pm0.37$  & \aer{0.08}{+0.05}{-0.04}  & 1   & \ldots &    1.0 &    3.9 &    1.0 & 0.01 & 0.02    \\
    H~37A &     E   & $0.87\pm0.12$  &  $27.40\pm0.16$  & \aer{3.83}{+0.47}{-0.42} & 1   & 1.00   &    9.0 &   98.4 &    1.0 & 0.00 & 0.00    \\
    H~37B &     S   & $1.27\pm0.17$  &  $10.70\pm0.15$  & \aer{1.48}{+0.50}{-0.38} & 1   & \ldots &    1.9 &   21.2 &    0.9 & 0.01 & 0.06    \\
    H~37C &     S   & $0.14\pm0.06$  &  $2.74\pm0.46$  & \aer{1.18}{+0.53}{-0.35}  & 2   & 0.62   &    1.9 &   21.2 &    1.0 & 0.00 & 0.00    \\
    H~40A &     E  & $0.69\pm0.09$  &  $22.00\pm0.16$  & \aer{0.48}{+0.12}{-0.09}  & 1   & 1.00   &    1.7 &   18.4 &    0.9 & 0.00 & 0.00    \\
  H~40BCE &     S  & $3.16\pm0.41$  &  $20.50\pm0.14$  & \aer{1.81}{+0.34}{-0.27}  & 2   & 0.85   &    1.2 &   13.1 &    0.7 & 0.05 & 0.25    \\
    H~40D &    S0  & $3.65\pm0.48$  &  $9.88\pm0.14$  & \aer{1.15}{+0.16}{-0.14}   & 2   & 0.40   &    3.4 &   36.7 &    1.0 & 0.00 & 0.00    \\
    H~42A &     E  & $0.71\pm0.08$  &  $31.90\pm0.06$  & \aer{1.78}{+0.14}{-0.13}  & 1   & 1.00   &   13.7 &   64.2 &    0.8 & 0.00 & 0.00    \\
    H~42B &    S0  & $0.20\pm0.04$  &  $4.76\pm0.06$  & \aer{0.43}{+0.14}{-0.09}   & 3   & \ldots &    1.1 &    5.1 &    0.8 & 0.00 & 0.06    \\
    H~42C &     E  & $0.14\pm0.04$  &  $4.71\pm0.06$  & \aer{0.30}{+0.08}{-0.06}   & 2   & 0.71   &    4.7 &   22.0 &    0.8 & 0.00 & 0.00    \\
    H~42D &    S0  & $0.02\pm0.01$  &  $1.08\pm0.21$  & \aer{0.33}{+0.13}{-0.09}   & 2   & \ldots &    1.4 &    6.8 &    1.0 & 0.00 & 0.00    \\
    H~59A &     S  & $5.92\pm0.80$  &  $1.48\pm0.04$  & \aer{1.36}{+0.24}{-0.20}   & 1   & 1.00   &    1.8 &    8.6 &    0.8 & 0.24 & 0.63    \\
    H~59B &     E  & $0.03\pm0.00$  &  $0.94\pm0.17$  & \aer{0.46}{+0.13}{-0.09}   & 3   & 0.48   &    0.9 &    4.3 &    1.0 & 0.00 & 0.01    \\
    H~59C &     S  & $0.19\pm0.04$  &  $0.41\pm0.29$  & \aer{0.91}{+0.22}{-0.17}   & 2   & \ldots &    0.9 &    4.3 &    0.9 & 0.00 & 0.01    \\
    H~62A &    S0  & $0.63\pm0.08$  &  $16.60\pm0.07$  & \aer{3.70}{+0.13}{-0.11}  & 13  & 0.09   &    2.0 &   10.1 &    1.0 & 0.00 & 0.00    \\
    H~62B &    S0  & $0.28\pm0.04$  &  $9.21\pm0.07$  & \aer{0.64}{+0.05}{-0.04}   & 4   & 0.43   &    2.0 &   10.1 &    0.9 & 0.00 & 0.00    \\
    H~62C &    S0  & $0.14\pm0.03$  &  $3.57\pm0.06$  & \aer{0.83}{+0.10}{-0.07}   & 9   & 0.13   &    0.7 &    3.4 &    0.9 & 0.00 & 0.03    \\
    H~62D &    S0  & $0.06\pm0.02$  &  $1.18\pm0.23$  & \aer{0.27}{+0.07}{-0.06}   & 1   & \ldots &    2.1 &   10.4 &    0.9 & 0.00 & 0.00    \\
   H~79AD &     S  & $0.78\pm0.09$  &  $4.29\pm0.06$  & \aer{1.29}{+0.15}{-0.12}   & 4   & \ldots &    0.5 &    2.3 &    0.8 & 0.01 & 0.08    \\
   H~79BC &    S0  & $0.87\pm0.11$  &  $6.07\pm0.06$  & \aer{0.43}{+0.08}{-0.06}   & 3   & 0.35   &    1.2 &    5.4 &    0.8 & 0.00 & 0.07    \\
   H~90BD &     S  & $1.09\pm0.14$  &  $12.70\pm0.02$  & \aer{0.93}{+0.09}{-0.07}  & 14  & 0.13   &    0.9 &    1.2 &    0.9 & 0.04 & 0.15    \\
    H~90C &     E  & $0.11\pm0.01$  &  $4.52\pm0.02$  & \aer{0.59}{+0.09}{-0.07}   & 10  & 0.20   &    0.9 &    1.2 &    1.0 & 0.00 & 0.20    \\
    H~92B &     S  & $1.73\pm0.19$  &  $14.50\pm0.18$  & \aer{1.61}{+0.18}{-0.13}  & 12  & 0.15   &    0.5 &    4.8 &    1.0 & 0.03 & 0.15    \\
    H~92C &     S  & $5.19\pm0.69$  &  $8.82\pm0.07$  & \aer{213.00}{+4.01}{-3.01} & 6   & 0.99   &    0.6 &    5.8 &    1.0 & 0.00 & 0.00    \\
    H~92D &     E  & $1.56\pm0.19$  &  $17.60\pm0.18$  & \aer{8.15}{+0.35}{-0.32}  & 6   & 0.08   &    0.5 &    4.8 &    1.0 & 0.00 & 0.00    \\
    H~92E &     E  & $0.37\pm0.06$  &  $11.00\pm0.14$  & \aer{0.51}{+0.09}{-0.08}  & 2   & 0.79   &    0.7 &    6.7 &    0.9 & 0.00 & 0.04    \\
    H~97A &    S0  & $0.34\pm0.07$  &  $13.20\pm0.11$  & \aer{0.83}{+0.15}{-0.12}  & 2   & 0.49   &    2.2 &   22.9 &    0.9 & 0.00 & 0.00    \\
    H~97D &     E  & $0.36\pm0.07$  &  $12.00\pm0.11$  & \aer{2.01}{+0.17}{-0.16}  & 1   & 1.00   &    4.2 &   43.6 &    1.0 & 0.00 & 0.00    \\
   H~100A &     S  & $3.49\pm0.44$  &  $8.30\pm0.07$  & \aer{2.95}{+0.50}{-0.43}   & 2   & \ldots &    2.6 &   18.3 &    1.0 & 0.02 & 0.18    \\
   H~100B &     S  & $1.30\pm0.15$  &  $4.49\pm0.08$  & \aer{1.51}{+0.46}{-0.35}   & 1   & 1.00   &    2.0 &   13.8 &    1.0 & 0.02 & 0.09    \\
   R~17A &     E  & $0.66\pm0.09$  &  $32.50\pm0.10$  & \aer{10.80}{+0.40}{-0.40}  & 1   & 1.00   &   41.2 &  326.0 &    1.0 & 0.00 & 0.00    \\
   R~17B &     E  & $0.31\pm0.04$  &  $8.08\pm0.09$  & \aer{13.20}{+0.50}{-0.50}   & 1   & 1.00   &    4.6 &   36.3 &    1.0 & 0.00 & 0.00    \\
   R~17C &     E  & $0.19\pm0.04$  &  $3.98\pm0.13$  & \aer{0.41}{+0.09}{-0.07}    & 1   & 1.00   &    5.0 &   39.3 &    1.0 & 0.00 & 0.00    \\
\enddata
\tablecomments{Columns are (1) galaxy ID (H for HCG, R for RSCG); (2) morphology based on best-fitting SED template (see L15 for details); (3) star formation rate; (4) stellar mass in units of $10^{10}$ solar masses; (5) total \x\ luminosity from point sources within a galaxy region in units of $10^{40}$\lunits\ in the 0.5-8.0~keV \chandra\ band; (6) number of detected point sources; 
(7) if there is a central source, fraction of the total luminosity contributed by the central source; (8) flux limit in units of $10^{-15}$~\funits; (9) $0.5-8.0$~keV luminosity limit in units of $10^{38}$~\lunits; (10)
probability that a source detected within the galaxy boundaries belongs to the CG galaxy; (11) probability of measuring a total galaxy luminosity greater than or equal to the observed value, given the luminosity detection limit for the galaxy and the \citet{gilfanov2004b} or \citet{mineo2012} XRB XLFs;
(12) as (11), but also assuming a range in luminosity for individual point sources of two orders of magnitude to account for variability.
}
\end{deluxetable*}

These flux limits can tell us what is the number of background AGNs we
expect to see inside the regions of the CG galaxies.  The
\lq\lq $\log N - \log S$\rq\rq\ relation of \citet{cappelluti2009}
gives the number of detected \x\ point sources per angular
area as a function of flux, established over 2 deg$^2$ in the COSMOS
field. Using our flux limits and this function, 
we calculate how many background AGNs we would expect to detect
in total in our galaxy regions.  
A measure of the probability  
$P_{\rm CG}$ (column 10 in
\tr{tab-results}) that a source detected inside a galaxy region is not
a background AGN is obtained if we subtract the number of expected
AGNs from the number of observed sources and then divide by the number
of observed sources.
According to \tr{tab-results}, for
41 out of 47 galaxies (87\%), $P_{\rm CG}\ge 0.8$. For the remaining
six galaxies (13\%), three have $P_{\rm CG}=0.7$ and three $P_{\rm
  CG}\le 0.6$.  We are thus confident that the great majority of our
sources most likely belong to their associated CG galaxies
and, overall, background contamination should not be
a major concern for this work. In what
follows we always flag the six galaxies with $P_{\rm CG}<=0.7$ to
take into account this relatively higher probability of AGN
contamination for \x\ point sources in these galaxies.

\section{SFR and \mstar}\label{sec-sfrmstar}
We obtain SFR and \mstar\ values for our CG galaxies as explained in
detail in L15. 

Since the SFRs are of particular significance for the conclusions
of this paper, we present here the key aspects in their
derivation. The SFRs are calculated by combining UV (\swift/UVOT 2000\AA)
and IR (\spitzer/MIPS 24\micron) photometry,
performed inside galaxy boundaries, which are Kron-aperture based
(the same as in \scr{sec-wide}), ensuring that $>90\%$ of the galaxy light is
included.  
All regions have been visually checked, and any contaminating
flux from foreground stars or background galaxies has been excluded.
Because of the relatively quiescent nature of our galaxies
in the low-SFR regime, populations of evolved stars may
contribute significantly both in the UV and at 24\micron.
The UV and 24\micron\ photometry has been corrected for this effect
by following the method of \citet{ford2013}. These authors use
the UV-to-3.6\micron\ and 24\micron-to-3.6\micron\ ratios established
for elliptical galaxies \citep{leroy2008}, to remove 
the best estimate of UV or 24\micron\ flux
not originating from star formation. 

Since the UV is a direct tracer of emission from massive O and B
stellar populations, while the 24\micron\ traces such emission
reprocessed by dust, our SFRs effectively include a correction for
intrinsic extinction.  To convert luminosities to SFRs the
\citet{kennicutt1998} calibration is used in the UV, and the
\citet{rieke2009} calibration in the IR (appropriate for the
luminosity range of our galaxies), leading to the combined
SFR determination
\begin{eqnarray}\label{equ-kenniri}
{\rm SFR} (M_{\odot} {\rm yr}^{-1})&=&9.5\times10^{-44} L_{2000{\textup{\AA}}} ({\rm erg s^{-1}}) \nonumber\\
          &+& 2.0\times10^{-43} L_{\rm 24\mu m} ({\rm erg s^{-1}}) \, .
\end{eqnarray} 

As a check, this relation is consistent with the more recent results of
\citet{hao2011}, obtained entirely independently from those
of \citet{kennicutt1998} and \citet{rieke2009}.
These authors cross-calibrate several multi-wavelength
SFR estimators for star-forming galaxies, including near-UV+25\micron.
The latter combination, which corresponds closely to our wavebands,
leads to
\begin{eqnarray}\label{equ-hao}
{\rm SFR} (M_{\odot} {\rm yr}^{-1})&=&1.1\times10^{-43} L_{\rm 2000{\textup{\AA}}} ({\rm erg s^{-1}}) \nonumber\\
          &+& 2.4\times10^{-43} L_{\rm 25\mu m} ({\rm erg s^{-1}}) \, .
\end{eqnarray} 

Using relation (\ref{equ-hao}) instead of (\ref{equ-kenniri})
for our galaxies, would lead to an average fractional increase in SFR
of 13\%, the maximum increase being 15\%. 

For stellar masses, we follow
\citet{eskew2012}, 
who derive
\exi
M_{\ast}=10^{5.65} F_{3.6}^{2.85} F_{4.5}^{-1.85} (D/0.05)^2  \ \ ,
\exo
where $F_{3.6}$ and $F_{4.5}$ are \spitzer/IRAC 3.6 and 4.5\micron\
fluxes, and $D$ distance in Mpc.

\section{Results}

\tr{tab-results} presents star formation rates, stellar masses and
total point-source luminosities (\lx) for galaxies in our sample.
Other relevant quantities are also included
as explained earlier.
We now present these results within the context of \x\ scaling
relations obtained independently for non-CG environments.
A further overview of these results is provided in Tables~\ref{tab-latecorrel}
and \ref{tab-earlycorrel} for late and early-type galaxies,
respectively. 
In these tables galaxies which show an ``excess'', a ``deficit'', or
neither because they 
are within 
the
$\pm 1\sigma$ 
scatter
of a known, non-CG, correlation are labelled
\bcblue{E}, \bcred{D} or \bcgrey{n}, respectively. For each
galaxy we also indicate whether there is evidence for a 
\lq\lq strong\rq\rq\ interaction. 
We use two main criteria to classify a galaxy as strongly interacting:
First, as already mentioned, some Kron galaxy regions contain more
than one letter-designated system (e.g. HCG 31 ACE). Such multi-galaxy systems
show pronounced 
morphological distortions and tidal features, which are
the signatures of active interaction and merging.
Using the
classification scheme of \citet{surace1998} \citep[see
  also][]{veilleux2002}, such interactions are at least type II
(\lq\lq First contact\rq\rq), where there is clear overlap.
Second, for a subset of our data, the level of interaction has been
studied in detail by means of kinematic, Fabry-Perot, data, and
specific interaction indicators \citep{plana1998,mendes1998,mendes2003,amram2004,torres-flores2009,torres-flores2010,torres-flores2014}. 
These indicators include
a highly disturbed velocity field, central double nuclei,
double kinematic gas component (i.e. a disagreement between
both sides of the rotation curve), kinematic warping, gaseous vs.~stellar 
major-axis misalignment (hereafter, G/SMM), 
tidal tails, high IR luminosity and central 
activity. As discussed by \citet[][herafter M98]{mendes1998}, the first three are 
most definitely associated with mergers.

\subsection{\lx\ vs. SFR}\label{sec-lxsfr}

\begin{deluxetable*}{cccc}
\tablecolumns{4}
\tablewidth{0pc} 
\tablecaption{CGs and Late-type Galaxy Correlations \label{tab-latecorrel}}
\tablehead{
\\
\colhead{ID}
& \colhead{\lx-SFR}
& \colhead{Strong}
& \colhead{Comments}
\\
\colhead{}
&\colhead{}
&\colhead{interaction?}
&\colhead{}
\vspace{0.1cm}
\\
\hline
\colhead{(1)}
&\colhead{(2)}
&\colhead{(3)}
&\colhead{(4)}
}
\startdata
16B  & \bcblue{E} & \ck\ & Strong AGN; interacting with A; G/SMM (M98). Single nuclear source.\\
92C  & \bcblue{E} & \ck\ & Strong AGN; tidal tails (TF09).       \\ 
37C  & \bcblue{E} & \ck\ & Interacting with H37A.\\  
59C  & \bcblue{E} &      &                    \\
31B  & \bcblue{E} & \ck\ & Interacting with ACE. \\
7C   & \bcblue{E} & \ck\ & Nuclear ring.       \\
31G  & \bcblue{E} & \ck\ & Peculiar morphology.    \\
79AD & \bcblue{E} & \ck\ & Peculiar morphology.     \\
37B  & \bcblue{E} &      &                    \\
100B & \bcblue{E} & \ck\ & Several signs of kinematic disturbance \citep{plana2003}.\\
16A  & \bcblue{E} & \ck\ & Possible LINER/weak AGN \citep[T14,][]{osullivan2014a,osullivan2014b};\\
                       &&& interacting with H16B \citep[bridge of diffuse emission,][]{jeltema2008}. \\
7A   & \bcblue{E} & \ck\ & Disturbed velocity field (TF09). \\        
31ACE& \bcblue{E} & \ck\ & Wolf-Rayet starburst interacting complex \citep{lopez-sanchez2004}.\\
92B  & \bcblue{E} & \ck\ & Unresolved \x\ emission. Tidal tails (TF09).\\
90BD & \bcblue{E} & \ck\ & Unresolved \x\ emission. Kinematically disturbed \citep{plana1998}.\\
100A & \bcblue{E} &      &                    \\
31Q  & \bcgrey{n} &      & Most isolated HCG31 galaxy.\\
40BCE& \bcgrey{n} & \ck\ & Only two point sources. G/SMM in 40E \citep{torres-flores2010}.\\
22C  & \bcgrey{n} &      &                    \\
59A  & \bcgrey{n} &      & TO/LINER (T14).\\
\enddata
\tablecomments{Columns are: (1) Spiral/irregular CG galaxies listed from top to bottom in order of {\it decreasing excess} relative to the M12 correlation (galaxies are HCGs except for designations starting with an R which are RSCGs); (2) excess (\bcblue{E}), Deficit (\bcred{D}), or none (\bcgrey{n}) relative to the M12 correlation; (4) evidence of strong interaction.
}
\end{deluxetable*}

In \fr{fig_mineo} we plot \lx\ against SFR for late-type galaxies
in this sample, together with the M12 best fit and $\pm 1\sigma$ scatter, 
and the
comparison samples of Douna et al. (2015) and \citet{basu-zych2013} (see
Section 6). Four galaxies are within the correlation's
scatter, while the remaining 16 (80\%) show an ``excess'' (higher scatter).
Specifically:

\begin{enumerate}[leftmargin=*]
\item In order of decreasing ``excess'' the 16 ``extreme \lx''
galaxies are:
\begin{itemize}[leftmargin=*]
\item The two known {\it strong}
  \x\ AGNs HCG 16B \citep{osullivan2014a,osullivan2014b} and HCG
  92C (NGC~7319). The former is a single nuclear \x\ point source.
  As shown in T14 the CG environment is very
  rarely host to unambiguously strong nuclear AGN emission; for this
  reason we have not attempted to exclude nuclear point-source
  emission in this work. These are the only late types in our sample
  that are strong AGNs, clearly the primary cause
  of the observed excess. 
  HCG 16B is also strongly interacting with its neighbor 16A
  (classification LINER/weak AGN in T14), to which it is linked
  by a bridge of diffuse (\x\ and optical) emission \citep{jeltema2008}.  The
  Fabry-Perot analysis of M98 detects a highly disturbed
  velocity field in HCG 16B, kinematic warping and G/SMM. 
  Similar analysis for 92C (NGC~7319) also detects prominent tidal
  tails \citep[][hereafter TF09]{torres-flores2009}.

\item HCGs 37C and 59C, which are both
classified star-forming \citep[morphologies S? in][]{devaucouleurs1991}.
The first is strongly interacting with HCG 37A. There is no evidence
for interactions for 59C \citep{konstantopoulos2012}.

\item HCG 31B, interacting with
the merging ACE complex.

\item HCG 7C, which has a nuclear
ring.

\item HCG 31G, morphologically peculiar.

\item The pair HCG~79AD, morphologically
classified peculiar, with a peculiar velocity field and highly interacting
\citep{mendes2003}).

\item HCG~37B.

\item HCG 100B of morphology S? and very strongly interacting
with its neighbors as shown by \citet[][highly disturbed velocity field,
double nuclei, tidal tails]{plana2003}.
 
\item HCG 16A (candidate LINER or weak AGN in 
T14, tidal tails in M98, close to
an intergalactic \htwo\ region, \citet{werk2010}).

\item HCG~7A 
(highly disturbed velocity field, TF09).

\item HCG 31ACE (a highly interacting
merging complex classified a Wolf-Rayet
starburst).

\item HCG~92B (NGC~7318B), for which tidal tails were detected by TF09.
 
\item HCG 90BD (highly interacting \citep{plana1998}).

\item HCG 100A.

\end{itemize}
\item The four galaxies that are within the scatter of the M12 correlation
are:

\begin{itemize}[leftmargin=*]
\item HCG~31Q, the most isolated, lowest SFR and \lx\ galaxy in the otherwise
highly active HCG~31 complex.

\item HCG 40BCE, a three galaxy interacting
complex, although it shows no disturbed velocity field in \citet{torres-flores2010}.

\item HCG 22C, for which TF09 do not
find a highly disturbed velocity field (they do detect
G/SMM).
\item HCG 59A (no evidence of interactions, classified
TO/LINER in T14).

\end{itemize} 
 
\end{enumerate}

\begin{figure}
\vspace{-1cm}
\hspace{-2cm}
\includegraphics[scale=0.75,clip=true]{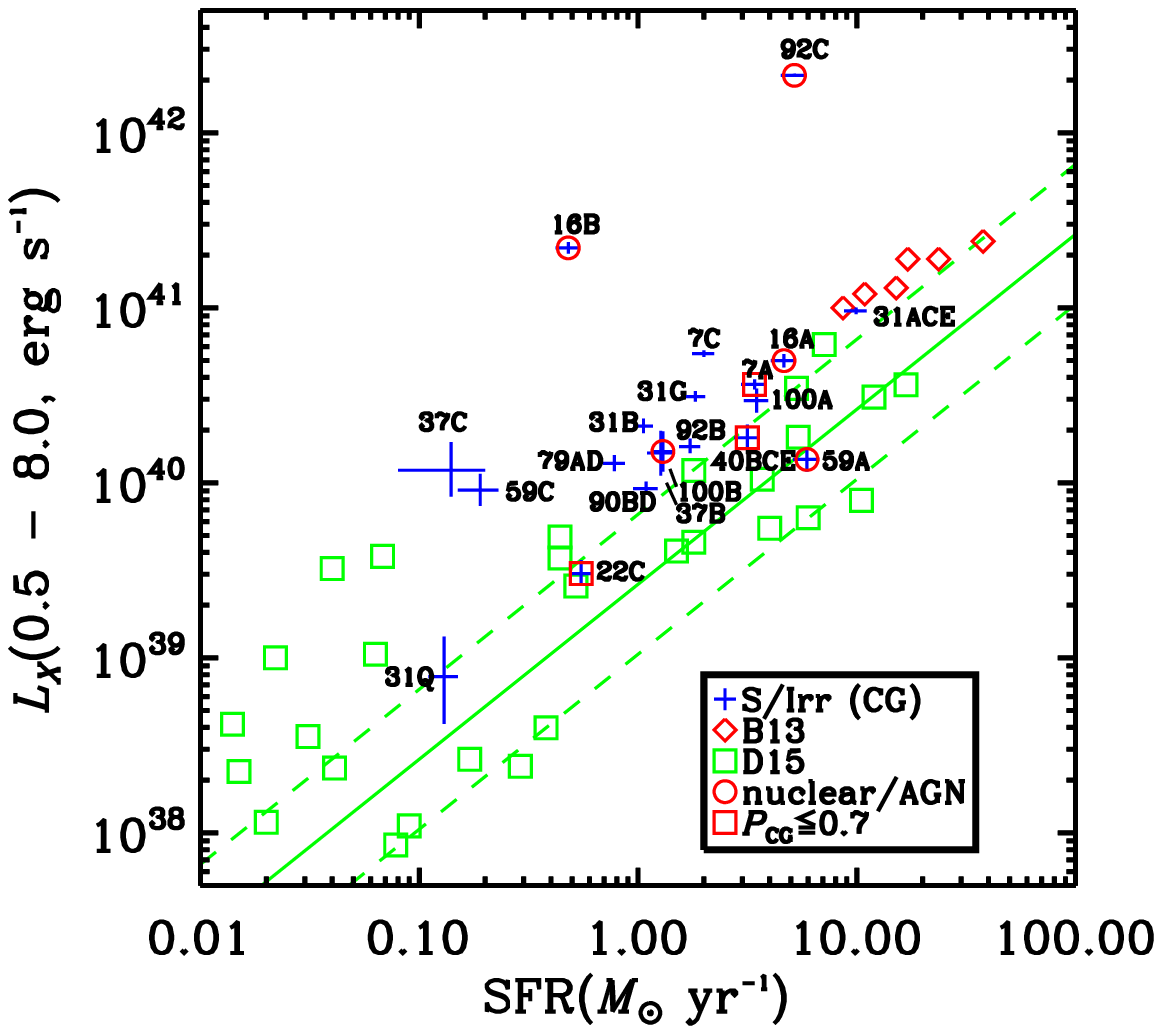}
\caption{Total \x\ luminosity from point sources, \lx, vs. SFR for late-type CG galaxies (blue symbols and error-bars). The five galaxies with a single, nuclear \x\ source and/or known independently to be
AGNs and/or having \lx~$\ge 10^{41}$~\lunits\ are flagged with red circles. The three late-type galaxies with detected sources having a lower probability of belonging to their galaxy (higher probability of being background AGNs) are flagged with red squares.
The solid (green) line is the M12 correlation for resolved HMXBs in nearby
star-forming galaxies. Its $\pm 1\sigma$ scatter is shown with dotted lines. 
The green squares are the Douna et al. (2015) data (D15) from nearby low-metallicity galaxies. The red diamonds are the Lyman-break galaxies of Basu-Zych et al. (2013, B13). Individual late-type HCG galaxies discussed in the text are indicated.
}
\label{fig_mineo}
\end{figure}

\subsection{\lx\ vs. \mstar}

\begin{figure}
\hspace{-2cm}
\includegraphics[scale=0.75,clip=true]{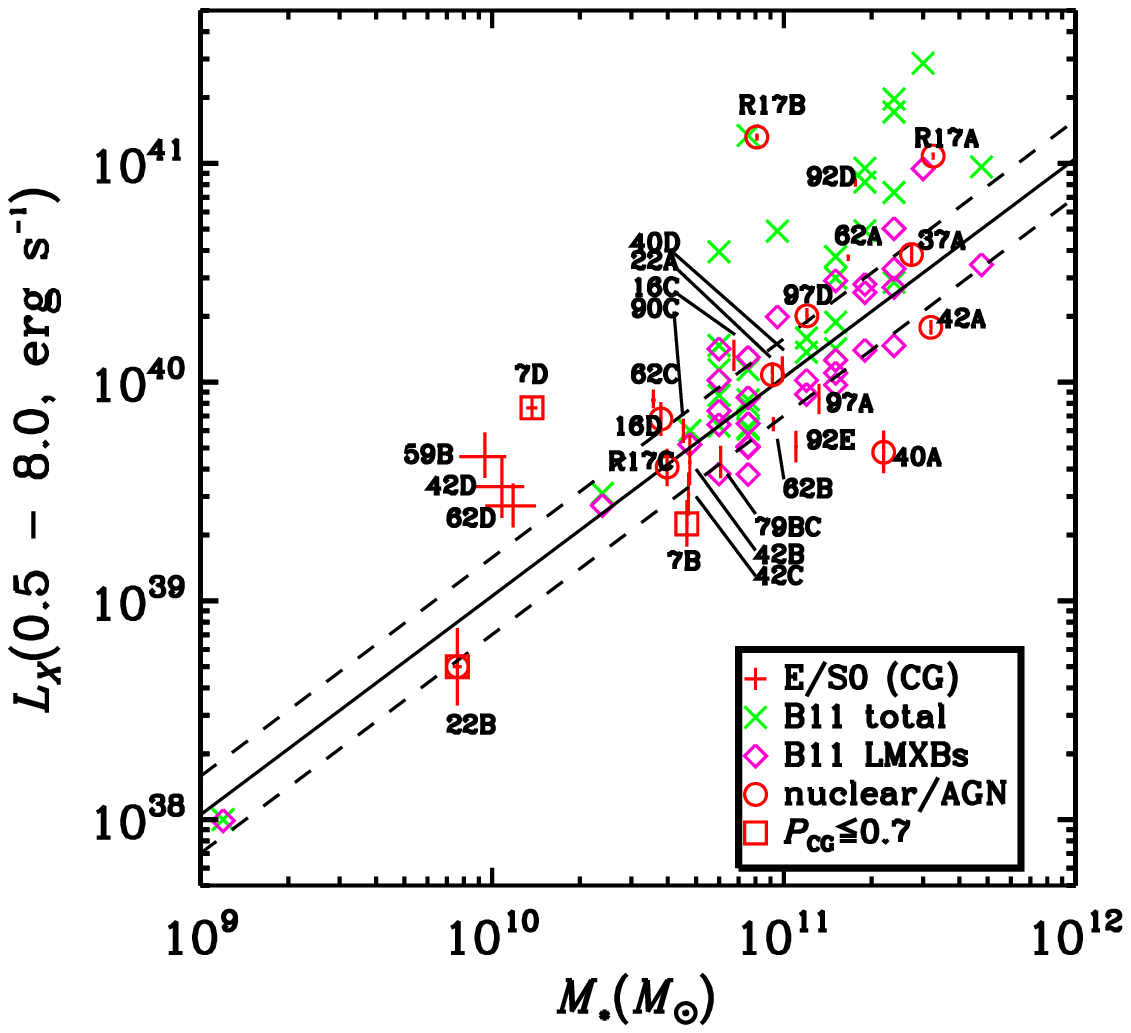}
\caption{Total \x\ luminosity from point sources, \lx, vs. \mstar\ for early type CG galaxies (red symbols
and error bars). The ten galaxies with a single, nuclear \x\ source and/or known independently to be
AGNs and/or having \lx~$\ge 10^{41}$~\lunits\ are flagged with red circles. The three galaxies with detected sources having a lower probability of belonging to their galaxy (higher probability of being background AGNs) are flagged with red squares. CG galaxies discussed in the text are labeled.
Labels for RSCGs begin with an R, otherwise galaxies are in HCGs.
The solid line is the correlation of B11
(with $\pm 1\sigma$ scatter dotted lines). Other symbols are:
{\it Green X's:} B11 sample total
(LMXB+AB+CV+AGN+hot gas) \lx\ values; 
{\it purple diamonds:}  B11 sample \lx\ values for LMXBs only.
}
\label{fig_bor}
\end{figure}

\begin{figure}
\hspace{-2cm}
\includegraphics[scale=0.75,clip=true]{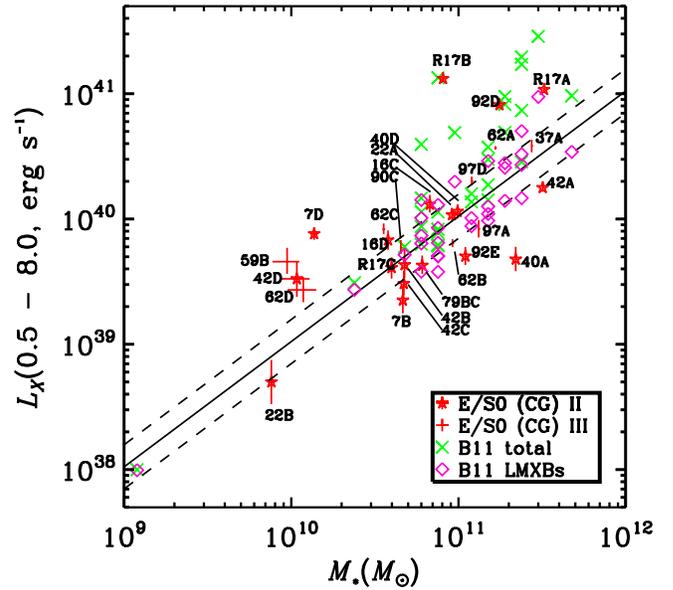}
\caption{Same as \fr{fig_bor}, but also
showing parent group \hone\ types:
Starred symbols are for CG galaxies 
in groups of evolutionary \hone\ type II; non-starred symbols are for galaxies
in groups of evolutionary \hone\ type III (see Table 1 for type definitions).
}
\label{fig_bor_h}
\end{figure}

In Figures \ref{fig_bor} and \ref{fig_bor_h} 
we plot \lx\ against \mstar\ results for early type galaxies,
together with the early-type sample of B11. Here, each
purple diamond point represents the total \x\ luminosity due to LMXBs
for a B11 galaxy. The best-fit and $\pm1\sigma$ scatter are shown by
the solid and dotted black lines.  The corresponding quantity,
\lx, for CG early-types is plotted with red error bars in
\fr{fig_bor}, which in \fr{fig_bor_h} are either starred or unstarred,
for galaxies of {\it group} evolutionary types II and III,
respectively.
Finally, in both figures, the green crosses show the
total \x\ luminosity taking into account
all types of \x\ emission (i.e. not just
LMXBs) in B11 galaxies. 
It appears that about
50\%\ (13/27) of CG galaxies are within the scatter of the
correlation; the rest have higher scatter, both above and below
the best-fit line.

Specifically:

\begin{enumerate}[leftmargin=*]
\item Galaxies with higher scatter than the B11 correlation and
on the high-\lx\ side (``excess'')  include:
\begin{itemize}[leftmargin=*]
\item The interacting pair RSCG~17B and A show the first and fifth highest
excess, respectively. For both galaxies, the only \x\ sources
are nuclear and dominate the
  \x\ luminosity at 1.1 and \ten{1.3}{41}\lunits, respectively,
  making them both candidates for AGN activity.

\item Second comes HCG 7D, with high levels of asymmetric emission both
  in the UV and in $R$
  \citep{konstantopoulos2010}. TF09 identify a
  strong G/SMM. However,
  background AGN contamination is relatively likely for this galaxy
  (\pcg~=~0.6). It is
  followed closely by HCG 59B for which in a deep co-added $B+R$ image
  \citet{konstantopoulos2012} detect a string of \htwo\ regions to the
  west of the galaxy, as well as a tidal bridge to neighbor A.  

\item HCG 92D (NGC~7318A) is located in a region of Stephan's Quintet which shows several star forming clumps and high-levels of diffuse emission. It shows
prominent tidal tails (TF09). 

\item HCG~42D is a low-\lx\ galaxy, that doesn't show any signs of 
strong interactions \citep{konstantopoulos2013}.

\item HCGs 62A, and C are closely interacting: A is optically
classified TO/LINER? with a radio excess from the radio-SFR
correlation (T14). HCG~62D
has no obvious interactions but it does show an excess of about the
same level as A and C.

\item Finally, HCG~16C, 
  shows several signs of strong
  interactions (M98): it has a highly disturbed velocity field,
  central double nuclei, G/SMM,
  a double kinematic gas component
  and kinematic warping. It is classified TO/SF in T14.
\end{itemize}

\item Galaxies within the correlation's $\pm 1\sigma$ scatter include:

\begin{itemize}[leftmargin=*]
\item HCG 16D is interacting with neighbor C, with whom it shares
some signs of strong interactions (M98) such as peculiar velocity
field and double nuclei. 
This is a single nuclear \x\ source, but
the galaxy is optically classified star-forming (T14).
\item HCG 37A, closely interacting with late-type 37C, has
a double kinematic gas component \citep{torres-flores2010}.
It also has a single nuclear
\x\ source and no independent nuclear classification.
\item HCG 40D, interacting with 40A.
\item The interacting pair HCG 79BC. The detailed work of \citet{durbala2008}, however,
finds no evidence of major mergers\footnote{The Fabry-Perot results of
these authors are not presented in a way that is directly comparable to those
for the rest of the systems.}.
\item HCG 62B, interacting with 62A.
\item RSCG 17C
and HCGs 22A, 22B, 42B, 42C, 90C,
97A, 97D are not obviously interacting with any neighbors.
RSCG 17C, HCGs 22B, 97D are single nuclear sources. HCG 22A
is classified AGN (T14).
Where available (TF09 for HCG 22A, B; \citet{plana1998} for HCG 90C)
Fabry-Perot kinematic results confirm this: only HCG 90C
shows one indicator for interaction, namely G/SMM. 
\end{itemize}

\item Galaxies that fall below the $\pm 1\sigma$ scatter bound include:
\begin{itemize}[leftmargin=*]
\item HCG 42A, a morphologically smooth elliptical.
Single nuclear source and optical classification ``non-SF''.
\item HCG 7B (no obvious interaction; corroborated by kinematic information, TF09). Background AGN contamination is likely (\pcg~=~0.5).
\item HCG 92E (NGC~7317), a rare case of a galaxy in Stephan's Quintet away from the \lq\lq main action\rq\rq. 
Only one minor kinematic interaction index in TF09.
\item HCG 40A, interacting with 40D, although not morphologically disturbed. 
Single nuclear source.
\end{itemize}

\end{enumerate}

We note that, as suggested by \fr{fig_bor_h}, there is no indication
that the group evolutionary type is a predictor of the location of an
early-type
CG galaxy in \lx-\mstar\ space.

\begin{deluxetable*}{cccc}
\tablecolumns{4}
\tablewidth{0pc} 
\tablecaption{CGs and Early-type Galaxy Correlations \label{tab-earlycorrel}}
\tablehead{
\\
\colhead{ID}
& \colhead{\lx-\mstar}
& \colhead{Strong}
& \colhead{Comments}
\\
\colhead{}
& \colhead{}
& \colhead{interaction?}
& \colhead{}
\vspace{0.1cm}
\\
\hline
\colhead{(1)}
&\colhead{(2)}
&\colhead{(3)}
&\colhead{(4)}
}
\startdata
R17B & \bcblue{E} & \ck\ & Candidate AGN.\\
7D  & \bcblue{E} & \ck\ & G/SMM (TF09). Possible background AGN contamination. \\
59B & \bcblue{E} & \ck\ & Western string of \htwo\ regions; tidal bridge to A \citep{konstantopoulos2012}. \\
92D & \bcblue{E} & \ck\ & Prominent tidal tails (TF09) and diffuse emission. Substantial amounts of hot gas due to \mstar~$>10^{11}$\msun (B11).\\
R17A & \bcblue{E} & \ck\ & Candidate AGN. Substantial amounts of hot gas due to \mstar~$>10^{11}$\msun (B11).\\
42D & \bcblue{E} &      & Low \lx; no signs of interactions.\\
62C & \bcblue{E} & \ck\ & Interacting with A.\\ 
62D & \bcblue{E} &      & \\
62A & \bcblue{E} & \ck\ & Interacting with C. TO/LINER? and radio excess (T14). \\
    &            &      & Substantial amounts of hot gas due to \mstar~$>10^{11}$\msun (B11).\\ 
16C & \bcblue{E} & \ck\ & Several kinematic signs of strong interactions (M98); TO/SF in T14.\\ 
16D & \bcgrey{n} & \ck\ & Interacting with C. Some kinematic interaction indicators (M98). \\
    &            &      & Single nuclear \x\ source, classification SF (T14).\\
97D & \bcgrey{n} &      & \\
37A & \bcgrey{n} & \ck\ & Interacting with C. Double kinematic gas component \citep{torres-flores2010}. Single nuclear \x\ source.\\
90C & \bcgrey{n} &      & \\
22A & \bcgrey{n} &      & AGN optical classification in T14.\\
40D & \bcgrey{n} & \ck\ & Interacting with 40A. G/SMM (TF10). \\
R17C & \bcgrey{n} &      & \\
42B & \bcgrey{n} &      & \\
79BC& \bcgrey{n} & \ck\ & Interacting pair.\\
62B & \bcgrey{n} & \ck\ & Interacting with 62A.\\
22B & \bcgrey{n} &      & \\
42C & \bcgrey{n} &      & \\
97A & \bcgrey{n} &      & \\
42A & \bcred{D}  &      & Single nuclear \x\ source. Classification ``non-SF'' (T14).\\
7B  & \bcred{D}  &      & Likely background AGN contamination.\\
92E & \bcred{D}  &      & \\
40A & \bcred{D}  &      & Interacting with 40D. Single nuclear \x\ source.\\
\enddata
\tablecomments{Columns are: (1) Elliptical/S0 CG galaxies listed from top to bottom in order of {\it decreasing excess} relative to the B11 correlation (galaxies are HCGs except for designations starting with an R which are RSCGs); (2) excess (\bcblue{E}), Deficit (\bcred{D}), or none (\bcgrey{n}) relative to the B11 correlation; (3) interaction level from kinematical work in the literature or
qualitative interaction indication (\ck).
}
\end{deluxetable*}

\section{Discussion}\label{sec-disc}
Our results suggest that X-ray point sources in CG galaxies show
trends that depend on galaxy morphology. Thus,
most late-type galaxies show \lx\ values higher than
expected from the M12 correlation and its scatter; the rest
appear to be close to the best fit line, with no lower-\lx\ scatter.
In contrast, 
early-type galaxies are consistent with the B11 correlation but
show larger scatter.

It is well established that many of the CG galaxies in this sample are
subject to strong interactions with neighbors. The evidence summarized
in Tables~\ref{tab-latecorrel} and \ref{tab-earlycorrel} suggests that
there might exist a connection between excess \lx, as compared to that
expected from non-CG environments, and interaction level. However,
whatever the physical mechanism behind the observed excesses might be,
it is not clear from these data that a high interaction level is
either necessary or sufficient to produce \lx\ excesses. There
is an {\it indication of a trend} for excess \lx\ to accompany high interaction
levels.  This is best seen in the case of the excess observed for
late-type systems relative to the \lx-SFR relation: Although 13 out of
16 such systems (81\%) showing an excess are also strongly interacting
(this includes the AGNs),
three systems show an excess without clear signs of interactions.
Also, one system is strongly interacting but does not show an excess.
This may simply be evidence for scatter in the relations combined with
small number statistics. 

We note that this connection can be investigated further by
quantifying interaction level, e.g. according to a scheme that weighs
interaction indicators.
Unfortunately, for the present sample there are too few
systems that have detailed and conclusive kinematic information for one
to carry out such an analysis.

Can the observed deviations be due to systematics? For instance,
it can be seen from \fr{fig_mineo} than
an increase by an order of magnitude in SFR would bring late type
systems into general agreement with the M12
correlation. 
The derivation of our SFRs is explained in \scr{sec-sfrmstar}.
As explained there, a recent alternative calibration might at worse increase
these values by up to $\sim 13\%$, on average. This is not nearly
sufficient to compensate for systems that show excess \lx\ in \fr{fig_mineo}.
In addition, for a subset of our systems,
\citet{bitsakis2014} have obtained
independent estimates using {\it Herschel} data and
UV-to-IR SED fitting. Their SFRs are systematically {\it lower} than ours, 
which would make our late-type excesses even more severe. 
There is thus no evidence that our SFRs are systematically too low,
which would lead to an artificial excess in the \lx--SFR relation.

Since our \x\ luminosities are point-source based, it is unlikely
that \lx\ is overestimated 
due to contamination from diffuse emission. 
It also seems improbable that, overall, background or nuclear AGN contamination
can significantly affect our results.
In total, there
are 20 late- and 27 early-types in our CG sample.
In Figs.~\ref{fig_mineo} and \ref{fig_bor} we flag
galaxies with a known AGN, a single nuclear
source or \lx~$\ge 10^{41}$~\lunits\ with solid red circles. We also flag galaxies with low
\pcg\ values with red squares. 
Conservatively excluding all these
galaxies (eight late-type and twelve early-type), still leaves 
11 out of 12 late-type galaxies (92\%) with an \lx\ excess,
and 8 out of 15 early-type galaxies (53\%) with an \lx\ excess or
deficit. Assuming the M12 and \citet{gilfanov2004b}
XRB XLFs, \pba\ is less than 0.05 for all but three of these, so stochastic
effects do not seem important. On the other hand,
as the \pbb\ values show, 
if a variability estimate is included, such effects may be more significant
for up to $\sim 45\%$ of galaxies.
Since many of these galaxies contain nuclear as well as non-nuclear sources,
some AGN contamination intrinsic to the galaxies cannot be excluded. 
The fact that we do not carry out
detailed spectral fitting should not affect the actual \lx\ values.
As shown by T14, even if individual $\Gamma$ values are incorrect,
fitted \nh\ and $\Gamma$ values work together to produce correct
\x\ luminosities. 

In this work we make the assumption that \x\ point sources in early-
and late-type galaxies are LMXBs and HMXBs respectively.  In reality,
however, it is possible that the \lx\ value for a galaxy region
includes a contribution from XRBs of the ``wrong'' type, i.e. either
LMXBs in early-types or HMXBs in late-types \citep[see][for an
  example]{luo2012}.  Since in most cases we only have very few,
low-luminosity sources for each galaxy region, this can make a
significant difference. Without high spatial resolution
multi-wavelength information for each point source we are not able to
separate LMXBs from HMXBs. We can however obtain a rough
estimate of the total contribution of LMXBs and HMXBs by using the
correlations of M12 and B11. We use the M12 correlation and SFRs for
early-type galaxies to obtain an estimate for the HMXB contribution
in these systems. Similarly, the B11 correlation and \mstar\ in 
late-type galaxies provide an estimate for the LMXB contribution.

\begin{figure}
\hspace{-2cm}
\includegraphics[scale=0.75,clip=true]{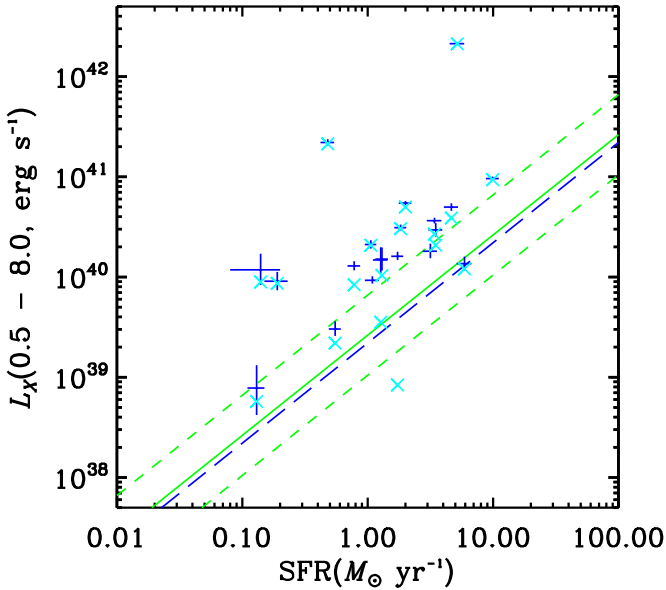}
\caption{Same as \fr{fig_mineo} but including
effect of subtracting an LMXB contribution, estimated from \mstar. The resulting
lower \lx\ values are shown by light-blue X's.}
\label{fig_mineo_bor}
\end{figure}

\begin{figure}
\hspace{-2cm}
\includegraphics[scale=0.75,clip=true]{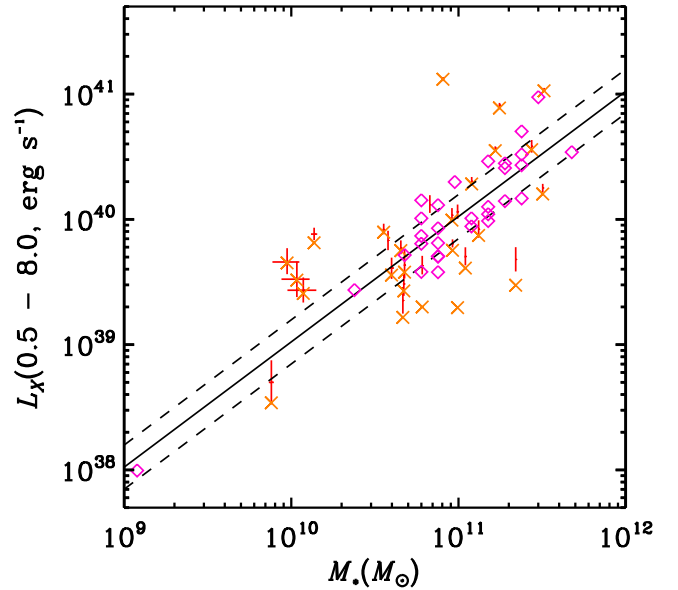}
\caption{Same as \fr{fig_bor} but showing the effect of subtracting a
  contribution from HMXBs, estimated from the SFR. The resulting lower
  \lx\ values are shown by orange X's.  }
\label{fig_bor_min}
\end{figure}

The X's in Figs~\ref{fig_mineo_bor} and \ref{fig_bor_min} show the
effect of subtracting these estimated contributions for late- and
early-type galaxies, respectively. It can be seen that the effect may
be significant only for the lowest-luminosity systems. Excesses from
the correlations can thus in general not be explained in this way.

For massive ellipticals (\mstar~$>10^{11}$\msun) B11 further caution
that the contribution from hot gas can be significant. If we include
S0 types, ten of our early-type galaxies have masses in this range.
Since not all show an excess, 
such a contribution may be relevant to the high \lx\ values for
HCG~62A, HCG~92D (NGC~7318A) and RSCG~17A. It is unlikely that
this is a very significant effect, since we are only including
emission from point sources.

The \x\ luminosity function of XRBs in
CGs is unknown and we have not carried out any incompleteness
corrections at the faint end. Significant numbers of undetected and/or
unresolved XRBs may further increase excess or decrease deficits. 

After taking into account possible AGN contamination (local or background),
a number of late-type CG galaxies still show unusually
high \lx\ for their SFRs.
It is well known that XRB numbers and luminosities are affected by
metallicity \citep{mapelli2009,zampieri2009}. XRB population synthesis modeling has shown a significant
increase in \lx/SFR with redshift \citep{fragos2013,tremmel2013,basu-zych2013} 
where metallicity is lower. Lower metallicities imply weaker stellar winds,
resulting in a lower mass loss of XRB primaries and, hence, an
increased production of massive black hole (BH) accretors. In turn,
BH-XRBs are more luminous than neutron-star (NS) XRBs for two reasons.
First, they can form stable Roche lobe overflow systems with high-mass
secondaries. Second, BHs have higher mass and higher Eddington limits,
thus giving rise to higher accretion rates. In this way, lower
metallicities can lead to higher luminosities both for
LMXBs and HMXBs.

In \fr{fig_mineo} the green squares indicate
the total \x\ luminosities from HMXBs in 
the nearby galaxy sample of \citet{douna2015}.
Each luminosity value represents the total
from individually detected HMXBs, i.e.~is equivalent to
our \lx\ values. These authors have specifically targeted 
low-metallicity galaxies, showing that there is an excess
in HMXB emission at low metallicities ($Z\lesssim 0.2$\zsun).
Further, the observed excess 
tends to increase with decreasing SFR.
We also show the local ($z<0.1$) ``Lyman Break Analogs'' of
\citet{basu-zych2013} with red diamonds. These authors find
an excess for their systems, which also have sub-solar metallicities
($Z\lesssim 0.6$\zsun). With the exception of our two highest excess systems
which are strong AGNs, it is clear that the rest of the
late-type CG galaxies occupy a similar locus in \lx-SFR space.
Similar to what noted by Douna et al. (2015)
there is also evidence for a stronger excess at lower SFR.
This is consistent with the reported low metallicities for
prominent CG star-forming systems such as HCG 31
\citep[$Z \sim 0.2$\zsun,][]{lopez-sanchez2004,gallagher2010} 
and HCG 92 \citep[$Z \sim 0.2,$\zsun][]{saracco1995,fedotov2011}.

There are currently no other metallicity measurements
for late type CG galaxies that would allow us to verify this low-metallicity
interpretation. Using the mass-metallicity relation of \citet{tremonti2004} leads
to estimates of solar metallicities, or higher, for these systems.
This, however, is at odds with existing direct measurements, e.g. for HCG~31.
CG late-type galaxies are however not typical star-forming galaxies:
As mentioned, most are undergoing strong interactions, likely leading to
significant gas and metal removal. As a result, although CG galaxies as a class
are not dwarfs, their mass-metallicity relation might be more similar
to that for dwarf galaxies \citep{kirby2013}. Formally applying 
such a relation consistently
produces low metallicities, in agreement with the few direct results
available at present.

Many of our individual \x\ point sources have luminosities greater
than the fiducial threshold of $10^{39}$\lunits\ and can thus be
classified as ``Ultra-luminous'' (ULXs). Independent evidence exists
for increased numbers of ULXs in low-metallicity environments
\citep{swartz2008,mapelli2010,kaaret2011,prestwich2013} consistent
with what we detect here.

Alternatively, and independently of a low-metallicity
interpretation, over an extended period in time interactions
can trigger and maintain elevated levels of star formation, and,
correspondingly relatively large populations of HMXBs. The \x\ 
luminosity registered at present for a given instantaneous SFR may then not
correspond to the one obtained in galaxies for which
strong interactions are not a prevailing characteristic, such as those
used to obtain the \lx-SFR correlation.

Note that any excess in {\it early-type} CG galaxies cannot be
explained as a metallicity effect. These systems are known to have high
metallicities and estimated ages
\citep{proctor2004,mendesdeoliveira2005}, even though
the co-existence of old ellipticals with young spirals
in some CGs suggests the possibility of replenishment with metal poor gas
from the intergalactic medium \citep{diaferio1994}.
Early-type systems likely simply show large scatter which is
a combination of local or background AGN contamination,
hot gas, as well as stochastic effects.

\section{Conclusions}
This paper represents the first study of the contribution of
individually detected \x\ point sources to the total
\x\ luminosity, and its correlation with star formation rate
and stellar mass,
in a sample of 15 compact groups (47 galaxies).  Our main results are:

\begin{enumerate}[leftmargin=*]

\item The probability, \pcg\ that \x\ point sources detected
within CG galaxy boundaries belong to the galaxy is high,
with 87\%\ of galaxies having \pcg~$\ge 0.8$. 

\item Late-type galaxies:

Excluding late-type galaxies that are known AGNs, have a single, nuclear
\x\ point source, \lx~$\ge 10^{41}$~\lunits, or have lower probability of hosting \x\ sources that are not background AGNs, 11 out of 12 galaxies (92\%) have an \x\ luminosity
which is higher than the one expected from the \citet{mineo2012} \lx\ - SFR 
relation and its
$1\sigma$ scatter. 
Excess luminosity appears to 
broadly
correlate with the level of interaction/merging activity, with
few exceptions.

\item Early-type galaxies: 

Excluding early-type galaxies that are known AGNs, have a single, nuclear
\x\ point source, \lx~$\ge 10^{41}$~\lunits, or have lower probability of hosting \x\ sources that are not background AGNs, 8 out of 15 galaxies (53\%) have an \x\ luminosity
that shows a larger scatter than the \citet{boroson2011}
\lx\ - \mstar\ relation.

\item Taking into account the luminosity detection limit for each galaxy, and
the XRB XLFs of \citet{mineo2012} and \citet{gilfanov2004b} for late- and early-type 
galaxies, the probability of obtaining stochastically \lx\ values
equal or higher to those observed is generally low, but increases
for about half of the galaxies if there is strong XRB variability.

\item For non-AGN late-type galaxies the excess \lx\ per SFR
in late-type systems is consistent with higher numbers of more luminous HMXBs
or ULXs in low-metallicity environments, 
as observed in nearby low-metallicity galaxies and local
Lyman Break Analogs.

\item The high \lx\ scatter for early-type systems likely
has a variety of origins, including
contamination from hot gas, background AGNs, local low-level AGN
activity, as well as stochastic 
and variability effects.

\end{enumerate}

Our results for late-type CG galaxies suggest that interactions can
have a dramatic effect on star formation and XRB populations. However,
not all interacting galaxies show excesses. This is likely related to
the amount and location of the gas reservoir, or perhaps a more recent
onset of interaction. Comparing XRB properties to the spatial distribution of
star formation and gas content can further explore the nature of these 
effects.

\acknowledgments A.H. and P.T. acknowledge funding provided through
\chandra\ Award No.~15620513 issued by the \chandra\ X-Ray
Observatory Center, which is operated by the Smithsonian Astrophysical
Observatory under NASA contract NAS8-03060.
We thank Andrew Ptak and Mihoko Yukita for useful discussions.  
We thank Bret Lehmer
for making his catalog of star forming galaxies available to us.
S.C.G., L.L. and T.D.D. thank the Natural Science and Engineering Research
Council of Canada and the Ontario Early Researcher Award Program for
support. This research has made use of data
obtained from the High Energy Astrophysics Science Archive Research
Center (HEASARC), provided by NASA's Goddard Space Flight Center.
This research has made use of the NASA/IPAC Extragalactic Database
(NED) which is operated by the Jet Propulsion Laboratory, California
Institute of Technology, under contract with the National Aeronautics
and Space Administration.

{\it Facilities:} \facility{\chandra, \swift, \spitzer}

\newcommand{\noopsort}[1]{}

\end{document}